\documentclass[%
 reprint,
 amsmath,amssymb,
 aps,
]{revtex4-2}

\usepackage{xcolor}
\usepackage{hyperref}
\hypersetup{colorlinks=true}

\hypersetup{
    colorlinks=true,  
    linkcolor=blue,   
    citecolor=red,  
    urlcolor=cyan,    
    filecolor=magenta 
}

\usepackage{url}

\usepackage{graphicx}
\usepackage{subfig} 
\usepackage{caption}
\usepackage[justification=justified, singlelinecheck=false]{caption}
\usepackage{subcaption}

\usepackage{graphicx}
\usepackage{dcolumn}
\usepackage{bm}

\begin{document}

\preprint{APS/123-QED}

\title{Identical Bosons, large occupation numbers and classical field description}

\author{Gaurav Goswami}
 \email{gaurav.goswami@ahduni.edu.in}
\affiliation{%
 Division of Mathematical and Physical Sciences, School of Arts and Sciences, Ahmedabad University,
 Navrangpura, Ahmedabad 380009, India
}%




\date{\today}

\begin{abstract}
For a system with a large number of identical Bosons, it is common to claim, often without any additional justifications, that, when the mean occupation number in a single particle state is sufficiently large, classical field description will be applicable. This is why e.g. for ultra-light dark matter, the classical field equations are used to compute its dynamics. 
In this work, we test the validity and robustness of this assumption based on the criterion $2 \sigma_\varphi < |\langle \varphi \rangle| $ for classical field behaviour and applying it to aribtrary quantum states. We find that an arbitrary state with large occupation number doesn't behave classically while imposing some restrictions on the state vectors can improve the classical behavior.
Since coherent states are known to have quasi-classical behaviour, we also ask how much deviation from coherent state can spoil the classical behaviour. Based on this analysis, 
we find that it is the proximity of the state to a large occupation coherent state rather than large occupation number itself which ensures validity of classical description. Implications of this for ultra light dark matter are discussed.

\end{abstract}

\maketitle







\section{Introduction and motivation}

Observations at multiple scales can be interpreted as evidence for the existence of Dark Matter (DM) \cite{Cirelli:2024ssz}.
Within this paradigm, we have some estimate of the mass density of DM at various length scales (e.g. the solar neighbourhood or at the scale of dwarf galaxies etc). 
However, since we don't know the particle mass for DM, we don't know DM number density. If DM consists of particles of a single species, for sufficiently small particle mass, the number density can be large enough that de Broglie wave packets of particles will start overlapping. In this regime, the mean number of particles per single particle state could vastly exceed unity - something of course possible only if the particles are identical Bosons \cite{Hui:2021tkt,Eberhardt:2025caq}.  

It is often assumed that, if the mean occupation number for a system of identical Bosons becomes large, it can be well described by classical field equations  - the regime of wave DM \cite{Hui:2021tkt,Eberhardt:2025caq}. This is why e.g. the simulations of wave DM often solve equations of classical field theory (see e.g. \cite{Schive:2025bcm} for review and \cite{Indjin:2025thr,Eberhardt:2025lbx} for some recent papers). Similarly, the experiments trying to directly detect wave DM (e.g. axions, ALPs etc) often rely on its temporal coherence properties \cite{OHare:2024nmr}, see also \cite{Cheong:2024ose}. 
Thus, in many such instances, the largeness of mean occupation number is automatically taken to imply the validity of classical field equations, without any additional justifications. When classical field description is known to be valid, it is possible to argue that the mean occupation number must be large \cite{Sakurai:1967advanced,LL:QED,Bialynicki-Birula:1977dab}, however, we know from e.g. quantum optics that light can have various quantum states only in some of which it behaves like a classical wave \cite{Navarrete-Benlloch:2022xpg}.  

The issue of validity of classical field description in the context of wave DM has been addressed previously but mostly from the perspective of whether interactions can cause the validity of classical approximation to break down and the time scale over which this could happen \cite{Sikivie:2016enz,Hertzberg:2016tal,Dvali:2017ruz,Arza:2019kab,Eberhardt:2023axk}.

Does the largeness of mean occupation number automatically imply the validity of classical field equations? Since improvements in observations are strongly constraining the particle mass of wave DM and such constraints are often based on the use of classical field equations - if the validity of classical field equations requires assumptions in addition to the largeness of mean occupation number, then, there is a possibility that 
small mass values of ultra light dark matter which are under tension may need to be revisited \cite{Proukakis:2023txk}, thus, this issue is closely connected to phenomenological questions.

Coherent states have a quasi-classical behaviour and the largeness of their mean occupation number does imply quasi classical behaviour \cite{Navarrete-Benlloch:2022xpg}. On the other hand,
the same is not true for Fock states. Thus, one might even ask: how much deviation from coherent states is okay to not ruin classical behaviour? 

The state of a quantum system of identical Bosons could be such that the classical field description may be valid. 
For laboratory systems such as BECs \cite{Leggett:2006-qls} and LASERs \cite{Navarrete-Benlloch:2022xpg}, an experimenter controls the conditions of the experiment in addition to preparing the state.
Since we don't know the details of the state and interactions of DM, one might wonder what happens for an arbitrary state in the state space of a system with a large number of identical Bosons.
Since the validity of classical approximation is assumed for any system with large mean occupation number, with no additional assumptions about its dynamics and state preparation, the underlying assumption seems to be that some reasonable fraction of randomly chosen states in the Hilbert space of a system of several identical Bosons have this feature. However, we will see that, classical field states are so special that, unless additional assumptions about the state preparation or dynamics of the system, which justify the use of classical field equations, are specified, automatically assuming that large mean occupation number implies the validity of classical field equations isn't justified. 

This paper is organised as follows:
in the next section, section \ref{sec:basics}, we begin by reminding the reader some fundamentals regarding the state space of the a non-relativistic system of multiple identical Bosons and then, in section \ref{sec:classicality}, we state our criterion for classical behaviour. In section \ref{sec:setup}, we describe our strategy for the analysis and then present the results that our strategy leads to in section \ref{sec:results}. The discussion section, section \ref{sec:discussion}, connects the analysis of the paper to broader issues in the literature.

\section{Classical field behaviour for a system of identical Bosons}

Let us think of the DM as consisting of a single species of particles, which we are going to assume are Bosons.
Thus, we think of a DM halo as a quantum system with a large number of non-relativistic identical Bosons.
If the state space of a single particle system is ${\cal E}$, the state space of two distinguishable particles is the tensor product space ${\cal E} \otimes {\cal E}$. The state space of a system of two identical Bosons is the subspace of ${\cal E} \otimes {\cal E}$ which is symmetric under the exchange of the two particles.
Note that, irrespective of the Hamiltonian of the system (in particular, irrespective of whether the particles are mutually interacting or not), a basis of the state space of two identical Bosons can be constructed from the basis of the single particle state space.
Similarly, the state space of a system with $N$ identical Bosons can be constructed - thus, the structure of state space is the same as that of tensor product of various spaces even in the presence of interactions.

\subsection{Structure of state space vs dynamics}
\label{sec:basics}

Such a system can be conveniently well described by using the formalism of quantized fields (see e.g. \cite{Pathria:1996hda} for a review of the same in the context of non-relativistic systems). 

Given a Bosonic quantum field ${\hat \varphi}({\bf r})$, corresponding to the observable ``total number of Bosons in the system", there is the Hermitian operator 
${\hat N}_{\rm tot} = \int d^3 {\bf r} ~ {\hat \varphi}^\dagger({\bf r}){\hat \varphi}({\bf r}) $. 
Consider an orthonormal basis $ \{ u_\beta ({\bf r}) \}$ in the space of square integrable functions over ${\mathbb R}^3$. The field operator can be expressed as a linear combination of the basis functions $ \{ u_\beta ({\bf r}) \}$ i.e. we can write
\begin{equation}
 {\hat \varphi}({\bf r}) = \sum_\beta {\hat a}_\beta  u_\beta ({\bf r}) \; .
\end{equation}
The Bosonic field commutation relations then imply that the quantities ${\hat a}_\beta$ satisfy commutation relations of the form 
$[{\hat a}_\beta, {\hat a}^\dagger_\gamma] = \delta_{\beta \gamma}$ etc. Additionally, the total number operator gets expressed as 
\begin{equation}
 {\hat N}_{\rm tot} = \sum_\beta {\hat a}^\dagger_\beta {\hat a}_\beta \; ,    
\end{equation}
and we can define mutually commuting ${\hat N}_{\beta} = {\hat a}^\dagger_\beta {\hat a}_\beta$ as the number operator corresponding to the mode with label $\beta$. It is interesting to recall that a function belonging to a basis such as $ \{ u_\beta ({\bf r}) \}$ also has an interpretation as the single-particle wave function. We can thus think of ${\hat N}_{\beta}$ as the particle number corresponding to the single-particle state $u_\beta ({\bf r})$. Since the various ${\hat N}_{\beta}$ commute with each other, they can be simultaneously diagonalised and there exists a particle-number basis of simultaneous eigenstate of all the operators ${\hat N}_{\beta}$. Since for each mode, commutation relations such as 
$[{\hat a}_\beta, {\hat N}_{\beta}] = {\hat a}_\beta$ etc hold good, the usual quantum harmonic oscillator machinery \cite{Cohen-Tannoudji:2019} can thus be borrowed and employed. 

Thus, the various modes behave like Harmonic oscillators - we haven't yet said anything about the Hamiltonian of the system or the interactions between the Bosons. Thus, it is reasonable to assume that the state space of a system with multiple identical Bosons is isomorphic to the state space of a collection of Bosonic oscillators irrespective of the interactions.
In the context of DM, we could additionally assume that the interactions between Bosons are such that the Hamilonian ${\hat H}$ of the quantum field commutes with ${\hat N}_{\rm tot}$. This will mean that (a) the total number of Bosons is conserved,
(b) the state space of the quantum field can be decomposed into direct sum of common eigenstates of ${\hat H}$ and ${\hat N}_{\rm tot}$.

Typically, the quantum field will have interactions - this will not change the state space, this can simply cause the particle-number eigenstates to not be stationary states. 
I.e., the interactions and other details can decide which states in this state space are stationary states and how they evolve in Schrodinger or interaction picture, but, the state at any time, will be a vector in the state space just mentioned.

In this work, we are going to be looking at a fixed mode but we will imagine that the number of Bosons in the mode could in principle vary due to interactions. Thus, states with indefinite number of Bosons will also be relevant for us. 
One might be concerned about whether quantum superposition of states with definite number of conserved particles could be non-existent due to some super-selection rule - we do not address this issue here.

\subsection{Criterion for classicality}
\label{sec:classicality}

What is the correct criterion for classicality in the context of wave DM? This is a subtle issue, not too different from the issue of definition of BEC (see e.g. chapter 2 of \cite{Leggett:2006-qls} for a discussion).

Let us begin by recalling that mean occupation number is defined to be the ratio of the number density of particles in phase space to the number density of quantum states in phase space. In many situations of interest only one or a few modes are such that their occupation numbers are large - this is obviously relevant in the context of BEC \cite{Leggett:2006-qls}, but, is also relevant in the context of wave DM \cite{Indjin:2025thr} -
thus, we will examine the case in which we have only one mode present.

If the quantum field is in a state in which it resembles a single mode of classical field, this mode of our interest must be in a state which shows the appropriate behaviour.
While there is no concept of field in QM of $N$ particles, in the equivalent quantum field formulation, there exists a field operator viz. ${\hat \varphi}({\bf x})$. In the Heisenberg picture, the Heisenberg equation of motion determines the time evolution of this field operator. For any state, one can then arrive at the corresponding Ehrenfest theorem by finding $\langle {\hat \varphi} (t, {\bf x}) \rangle$. This Ehrenfest theorem will imply that the expectation value of quantum field operator in any state $\langle \varphi \rangle$ will always satisfy the same equations that the field in classical field theory satisfies. 

If a quantum state is such that the standard deviation in field value computed in that state  i.e. $\sigma_{\varphi}$ is too small as compared to $| \langle \varphi \rangle |$, the mean value of the field reliaby captures the dynamics. Thus, such a state of the quantum field can be said to be a classical field state since classical equations of motion can be said to be satisfied
\footnote{We note that, since, classical field equations are used in simulating wave DM halos, their validity is more general than the validity of other criteria such as Penrose Onsager criterion for spatial coherence \cite{Indjin:2025thr}, which are tested for solutions of classical field equations.}
.
Notice that our requirement will need $\langle \varphi \rangle$ to be non-zero - thus, we could keep track of the quantity $\sigma_{\varphi} / | \langle \varphi \rangle |$ and check whether it is sufficiently small compared to 1. How small is enough? While, even a value smaller than 1/2 is interesting, for us, the benchmark case for comparison will always be that of coherent state (see \cite{Cohen-Tannoudji:2019} for a review). 

Before proceeding, we remind the reader that in a Fock state (i.e. eigenstate of number operator), $\langle \varphi \rangle = 0$, so, they aren't classical, no matter how large the occupation number.
Since $\sigma_\varphi$ has $\langle {\hat \varphi}^2 \rangle$, one might be concerned about the divergences in this quantity for arbitrary state (including vacuum state, coherent state etc) and hence the renormalization of this composite operator. We will not be concerned about this issue in this work - as some aspects of dealing with it will depend on the details of the interactions and observed values of some quantities.

\section{Details of numerical work}

\subsection{Basic set up}
\label{sec:setup}

Will an arbitrarily chosen superposition of Fock states with large mean occupation number be well described by classical field theory?
Since the state space of a single Bosonic oscillator is infinite dimensional and we have only a finite (albeit large) number of Bosons available, the state of the system of Bosons of interest would be a finite-dimensional vector in this state space. Let us call the dimension of the space to which our state vector belongs $N_{dim}$.

For an arbitrary normalized ($\sum_{n = 0}^{\infty} |c_n|^2 = 1$) state of the form $| \psi \rangle = \sum_{n = 0}^{\infty} c_n | n \rangle $, the mean occupation number is
$\langle N \rangle = \sum_{n = 0}^{\infty} n |c_n|^2$, while the expectation value of the field operator, $\langle \varphi \rangle = C_A ({\hat a} + {\hat a}^\dagger)$ is given by
\begin{equation} \label{eq:mean}
 \langle \hat \varphi \rangle = C_A \left[ \sum_{n=0}^{\infty} c_n^* \left( c_{n+1}\sqrt{n+1} + c_n \sqrt{n} \right) \right] \; ,
\end{equation}
where, $C_A$ is a constant, which we set to one for convenience. 
Similarly, the expectation value of square of the field is
\begin{eqnarray} \label{eq:var} 
 \langle \hat \varphi^2 \rangle 
 &=& C_A^2 \sum_{n = 0}^{\infty} 
 \left[
  \sqrt{(n+1)(n+2)}~ c_n^* c_{n+2} \right. \nonumber \\
&& \left. + \sqrt{2n + 1} ~|c_n|^2 
 + \sqrt{n(n-1)} c_n^* c_{n-2}
 \right]
   \; ,
\end{eqnarray}
note that in this expression, in the sum, for $n=0$ and $n=1$, the last term doesn't contribute due to the factor $\sqrt{n(n-1)}$, so, we don't need $c_{-1}$ and $c_{-2}$.
From $\langle \varphi^2 \rangle$ , we can find 
$\sigma_{\varphi} = \sqrt{\langle \varphi^2 \rangle - \langle \varphi \rangle^2 }$.

Before proceeding, we note that for a coherent state with parameter $\alpha$ (defined by ${\hat a} | \alpha \rangle = \alpha | \alpha \rangle$, see \cite{Cohen-Tannoudji:2019} for more details), 
\begin{equation} \label{eq:coh-coeff}
 c_n = \frac{\alpha^n}{\sqrt{n!}} e^{- \frac{|\alpha|^2}{2}} \; ,    
\end{equation}
This gives, $\langle N \rangle = |\alpha|^2$, $\langle \varphi \rangle \propto {\rm Re}(\alpha) = 2 C_A \langle N \rangle^{1/2}$ and 
one finds that, our measure of classicality is 
\begin{equation}
\frac{\sigma_{\varphi}}{\langle \varphi \rangle} = \frac{1}{2 |\alpha| } \; ,
\end{equation}
which would imply that when, $\langle N \rangle$ is increased, this ratio becomes small. Thus, a large $|\alpha|$ coherent state (corresponding to large occupation number $\langle N \rangle$) will resemble a classical state even by this criterion. 
In contrast, a Fock state has zero $\langle \varphi \rangle$ and so, a non-zero $\sigma_\varphi$ will never be small as compared to $\langle \varphi \rangle$, it is thus not classical by this criterion no matter what the occupation number is. 

In practice, instead of summing over the entire range in eq (\ref{eq:mean}) and eq (\ref{eq:var}), we sum over $n$ going from $0$ to a large positive integer $N_{dim}$ (mentioned above). Since for every $n$, we need coefficients such as $c_{n+1}$ and $c_{n+2}$, we choose to generate coefficients up to $n = N_{max} = N_{dim} + 2$. 
One can then consider various cases e.g. $N_{dim} = 25, 50, 100$ etc. 

The coefficients could be chosen randomly distributed in various ways as described below.
To generate the state vectors, we will take the expansion coefficients $c_n$s to be real numbers.
This can be justified by the fact that even expressions such as ${\hat a}| n \rangle = \sqrt{n}| n-1 \rangle$ can have some phases, these phases can be used to ensure that the coefficients $c_n$ are real.

For each state vector so generated, one obtains a unique value of $\langle N \rangle$, $\langle \varphi \rangle$, $\sigma_{\varphi}$ and hence $\sigma_{\varphi} / \langle \varphi \rangle$.
One could then do this for a large number ($N_{\rm sample}$) of state vectors and examine what happens.
One can of course vary $N_{\rm sample}$, the number of vectors in the sample of interest e.g. $10^4$, $2 \times 10^5$ etc.

\subsection{Results for various cases}
\label{sec:results}

\subsubsection{Case I}

Let us first examine the case in which all $c_n$s are uniformly distributed pseudo-random numbers within the interval $(-1,1)$. 
Once a pseudo random vector is obtained, we normalise it and apply the expressions given above to compute the quantities $\langle N \rangle$, $\langle \varphi \rangle$, $\sigma_\varphi$, $\sigma_{\varphi} / \langle \varphi \rangle $ etc for each state vector in our sample. 

\begin{figure*}[t]
\centering
    \subfloat[]{%
        \includegraphics[width=0.45\textwidth]{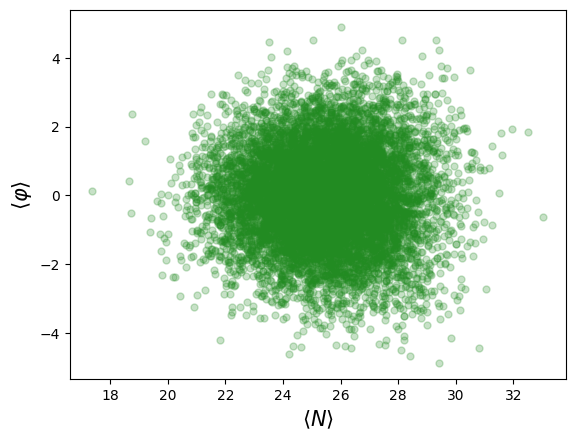}%
        \label{fig1:sub1}%
    }%
    \hfill
    \subfloat[]{%
        \includegraphics[width=0.45\textwidth]{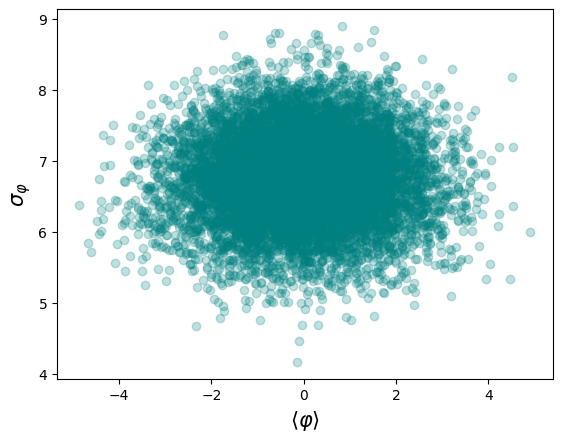}%
        \label{fig1:sub2}%
    }%
    
    \subfloat[]{%
        \includegraphics[width=0.45\textwidth]{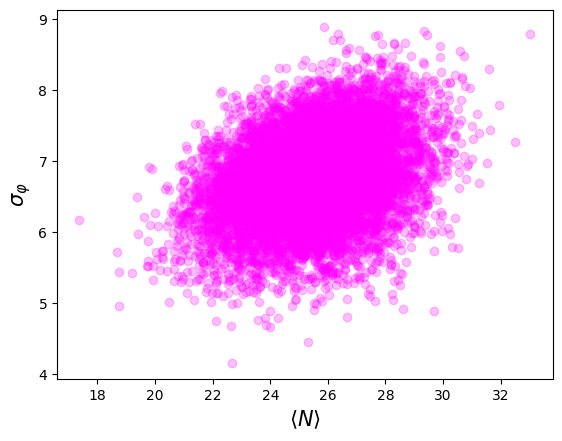}%
        \label{fig1:sub3}%
    }%
    \hfill
    \subfloat[]{%
        \includegraphics[width=0.45\textwidth]{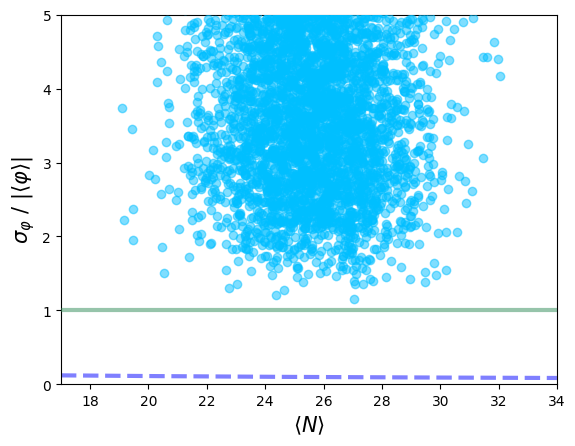}%
        \label{fig1:sub4}%
    }%
    
    \caption{When the coefficients $c_n$ are uniformly distributed within $(-1,1)$, there is no correlation between the most quantities of our interest (except between $\sigma_\varphi$ and $\langle N \rangle$ in \protect\subref{fig1:sub3}, with Pearson coefficient = 0.31 i.e. a weak correlation). In \protect\subref{fig1:sub4}, the dashed line corresponds to $\sigma_{\varphi} / | \langle \varphi \rangle|$ for $c_n$s for a coherent state. }
    \label{fig:posneg}
\end{figure*}

For the case with $N_{dim} = 50$ and $N_{\rm sample} = 10^4$, following this procedure gives us fig (\ref{fig:posneg}). It is worth noting that for $N_{dim} = 50$, the typical vectors generated have their $\langle N \rangle$ approximately around 25. 
While it is possible to have this large mean occupation number, $\langle \varphi \rangle$ can be large or small - as fig \ref{fig:posneg} \protect\subref{fig1:sub1} shows. 
In other words, what we don't see in this figure is important: randomly chosen vectors in state space with large 
$\langle N \rangle$ do not necessarily have larger 
$\langle \varphi \rangle$ and smaller $\sigma_\varphi$. Thus, large occupation number doesn't imply a state resembling classical behavior.

For subfigure \protect\subref{fig1:sub4} of fig (\ref{fig:posneg}), most states have $\sigma_{\varphi} / | \langle \varphi \rangle|  \gg 1$ as most of the states have $\langle \varphi \rangle$ close to zero, we have only shown the behaviour near 1. 
Sub-figure \protect\subref{fig1:sub4} also shows that in this sample, there are no states such that $\sigma_\varphi < |\langle \varphi \rangle|$. In contrast, for $\langle N \rangle$ in this range, the coherent state gives $\sigma_{\varphi} / | \langle \varphi \rangle| $ of around 0.1 (dashed line).
Thus, under these conditions, the typical value of $\sigma_{\varphi} / |\langle \varphi \rangle|$ for a randomly chosen state is much greater than 1.

Another observation worth paying attention to is the lack of correlation between the various quantities of our interest, though there is a weak positive correlation between $\sigma_\varphi$ and $\langle N \rangle$, see the caption of fig (\ref{fig:posneg}).

\begin{figure*}[t]
\centering
    \subfloat[]{%
        \includegraphics[width=0.45\textwidth]{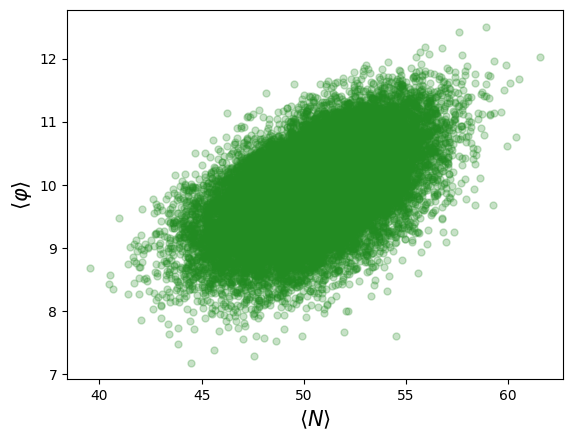}%
        \label{fig2:sub1}%
    }%
    \hfill
    \subfloat[]{%
        \includegraphics[width=0.45\textwidth]{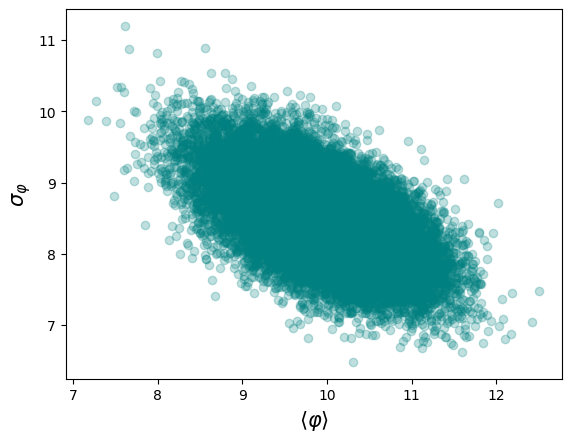}%
        \label{fig2:sub2}%
    }%
    
    \subfloat[]{%
        \includegraphics[width=0.45\textwidth]{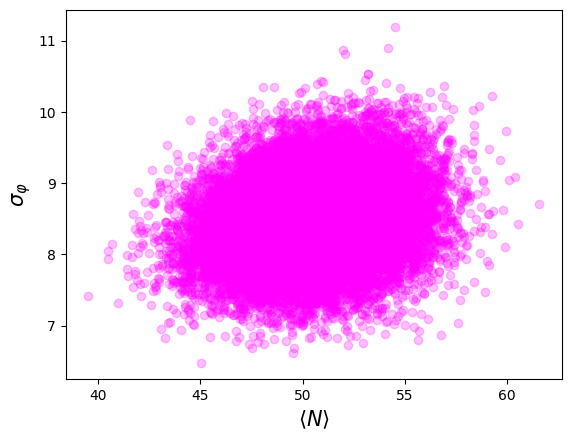}%
        \label{fig2:sub3}%
    }%
    \hfill
    \subfloat[]{%
        \includegraphics[width=0.45\textwidth]{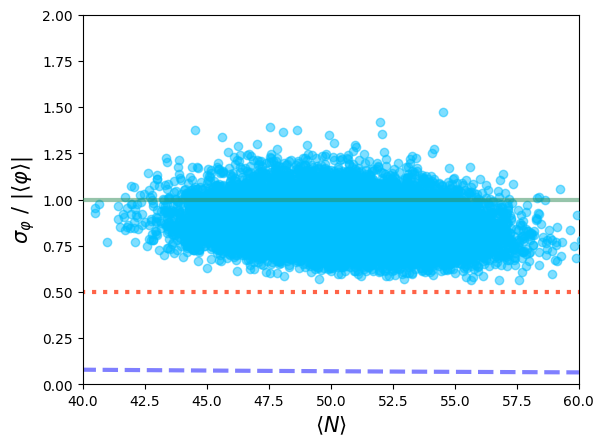}%
        \label{fig2:sub4}%
    }%
    
    \caption{When the coefficients $c_n$ are uniformly distributed within $(0,2)$, we begin to see weak correlations between the various quantities of our interest: the values of Pearson coefficient for the sub-figures being 0.55, -0.57, 0.20, -0.20 respectively. In \protect\subref{fig2:sub4}, in addition to the dashed line for the case of coherent state, the solid line and dotted line correspond to $\sigma_{\varphi} / | \langle \varphi \rangle| = 1$ and $1/2$.}
    \label{fig:pospos}
\end{figure*}

\subsubsection{Case II}

Could our results for the last case be sensitive to the value of $N_{dim}$? Could increasing the value of the number of vectors in the sample i.e. $N_{\rm sample}$ change anything substantial?
We verified that increasing $N_{dim}$ from 50 to 100 will give similar results, though the typical $\langle N \rangle$ of the vectors in the sample will now be around 50. Similarly, changing the number of vectors in the sample from $10^4$ to $10^5$ doesn't change these results i.e. the chances of randomly finding a state with $\sigma_{\varphi} / | \langle \varphi \rangle| < 1$ are less than 1 in $10^5$.

\subsubsection{Case III}

Let us now examine the case in which $c_n$s are all positive, in particular, suppose they are uniformly distributed pseudo-random numbers within the interval $(0,2)$. The results for this case are shown in fig (\ref{fig:pospos}) for $N_{dim} = 100$ and $N_{\rm sample} = 25000$. We see that merely insisting that all the coefficients $c_n$ be positive introduces correlations between various quantities. The expectation value of the field is now positive for all the states and $\langle \varphi \rangle$ increases with increasing $\langle N \rangle$ (see fig (\ref{fig:pospos}) caption for relevant details). Moreover, there is a moderate negative correlation between $\langle \varphi \rangle$ and $\sigma_\varphi$. This is very encouraging because for a classical state, we do need large $\langle \varphi \rangle$ and small $\sigma_\varphi$ - could this be achieved be increasing $\langle N \rangle$? 

Subfigure \protect\subref{fig2:sub4} of fig (\ref{fig:pospos}) shows that, $\sigma_{\varphi} / | \langle \varphi \rangle|$ is below 1 for a large fraction of cases, it is still not smaller than 1/2. Moreover, it is worth noting that while the negative correlation between $\langle N \rangle$ and $\sigma_{\varphi} / | \langle \varphi \rangle|$ is encouraging (in the sense that increasing $\langle N \rangle$ suggests a decrease in $\sigma_{\varphi} / | \langle \varphi \rangle|$), this correlation is very weak.
These observations are consistent with the constraint of positive-semidefiniteness correlation matrices.

We thus conclude that the among $2.5 \times 10^4$ states, it is possible to make $\sigma_{\varphi} / |\langle \varphi \rangle|$ smaller than 1 but not too small compared to 0.5. The chances of randomly being in a state with $\sigma_{\varphi} / |\langle \varphi \rangle|$ comparable to its value in coherent state are very small - the probability that a state chosen this way has $\sigma_{\varphi} / |\langle \varphi \rangle| < 1/2$ is definitely less than $2.5 \times 10^{-4}$. 

These examples suggest that a state resembling  classical field is possible provided the coefficients $c_n$ behave in the ``right" way.
Clearly, when the coefficients $c_n$ take the form valid for coherent states mentioned above, then, for large value of the modulus of coherent state parameter $|\alpha|$, the ratio $\sigma_{\varphi} / | \langle \varphi \rangle|$ is the smallest possible (without squeezing). 

\subsubsection{Case IV}

\begin{figure*}[t]
\centering
    \subfloat[]{%
        \includegraphics[width=0.45\textwidth]{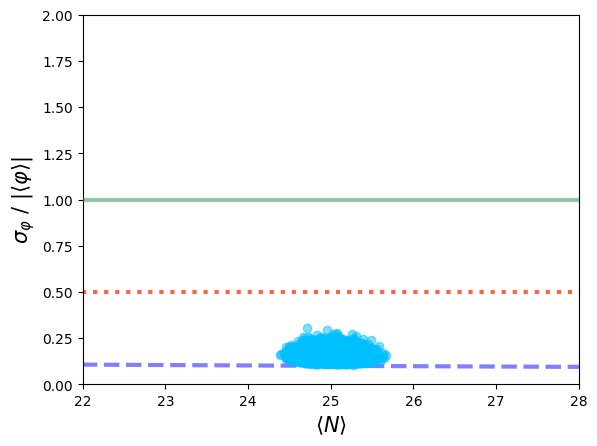}%
        \label{fig3:sub1}%
    }%
    \hfill
    \subfloat[]{%
        \includegraphics[width=0.45\textwidth]{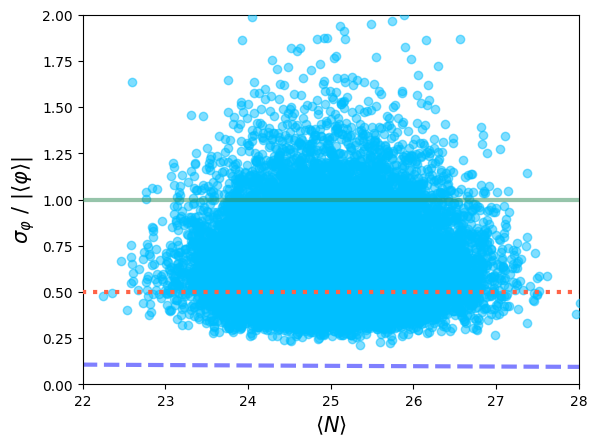}%
        \label{fig3:sub2}%
    }%
    \subfloat[]{%
        \includegraphics[width=0.45
        \textwidth]{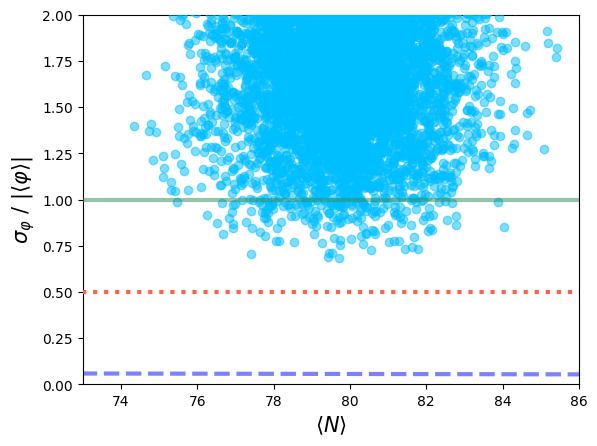}%
        \label{fig3:sub3}%
    }%
    \caption{Small and large deviations from coherent states: \protect\subref{fig3:sub1} $|\alpha|^2 = 25$, $f = 0.1$, \protect\subref{fig3:sub2}, 
    $|\alpha|^2 = 25$, $f = 0.5$, 
    \protect\subref{fig3:sub3} $|\alpha|^2 = 80$, $f = 2.0$. As we increase the deviation from coherent state by increasing $f$, the ratio $\sigma_{\varphi} / | \langle \varphi \rangle|$ becomes larger and eventually, classical behaviour is lost.
    }
    \label{fig:coherent}
\end{figure*}

Given this, it make sense to look at the vectors whose coefficients are similar to coherent states. We thus consider vectors whose expansion coefficients are normally distributed pseudo random numbers around $c_n$ values for coherent state. For each $n$, we will have a normal distribution with mean given by the expression in Eq (\ref{eq:coh-coeff}) and standard deviation $\sigma$ which is a factor $f$ times the mean value. We can then change $f$ and see how that affects the value of $\sigma_{\varphi} / | \langle \varphi \rangle| $ which can be achieved.

For very small $f$, $\langle N \rangle$ for the various states generated this way tends to have a very small spread around the mean value $|\alpha|^2$, with $\alpha$ being the coherent state parameter for the state around which the coefficients of our states are normal distributed. 
Since the coefficients $c_n$ for a vector are now distributed in a specific way, it shouldn't be surprising that some correlations appear between some of the quantities of our interest. For very small $f$, the Pearson correlation coefficient between $\langle \varphi \rangle$ and $\langle N \rangle$ tends to approach $+1$. 

The results of this exercise are shown in fig (\ref{fig:coherent}) which has been generated for $N_{dim} = 100$ and $N_{\rm sample} = 25000$ and which only shows the plots of 
$\sigma_{\varphi} / | \langle \varphi \rangle| $ against $\langle N \rangle$ for
varying $f$ and $\alpha$.
Firstly, when $f$ is sufficiently small, the results should be very close to the results in the case of coherent state - this is seen in \protect\subref{fig3:sub1} of the figure which corresponds to $f = 0.1$. Note how all the states in the sample have $\sigma_{\varphi} / | \langle \varphi \rangle| < 1/2$ and most have $\sigma_{\varphi} / | \langle \varphi \rangle| < 1/4$. This state, though it is not strictly coherent state, deviates so less that it appears classical.
We have verified that choosing an even smaller $f$ makes the scatter of points sit almost on top of the dashed blue line. 

As we increase $f$, the deviation of the state from coherent state increases (see e.g. \protect\subref{fig3:sub2} of fig (\ref{fig:coherent}), which is for $f = 0.5$) and we get a larger value of $\sigma_{\varphi} / | \langle \varphi \rangle| $ i.e. the deviations from classical field equations begin to increase. Still, in this case, a fair fraction (5513 out of $2.5 \times 10^4$, roughly 22\%) of states behave classically in the sense that $\sigma_{\varphi} / | \langle \varphi \rangle| < 1/2$.
For a run with $N_{dim} = 100$ and $N_{\rm sample} = 10^5$, $|\alpha|^2 = 25$ and $f = 1$, we find that only in 427 out of $10^5$ cases (i.e. roughly $0.4 \%$) is $\sigma_{\varphi} / | \langle \varphi \rangle| < 1/2$. 
On the other hand, when $f = 2$ corresponding to part \protect\subref{fig3:sub3} of fig (\ref{fig:coherent}), none of the states have $\sigma_{\varphi} / | \langle \varphi \rangle| < 1/2$. 
This case begins to resemble the case shown in
 subfigure \protect\subref{fig1:sub4} of figure \ref{fig:posneg}.
It is to be noted that for \protect\subref{fig3:sub3} of Fig (\ref{fig:coherent}) the coherent state parameter $\alpha$ is such that $\langle N \rangle = 80$.

The results of this case imply that even small deviations from coherent state take us into the non-classical regime. 
In particular, to keep $\sigma_\varphi / |\langle \varphi \rangle|$ smaller than 1/2, the coefficients $c_n$ have to be sufficiently close to their values for a coherent state. 

\section{Discussion}
\label{sec:discussion}

In this work, our criterion for classical field behaviour, as explained in section \ref{sec:classicality}, is that a state which is such that the standard deviation of Bosonic field operator is small (at least half) as compared to its expectation value shall be accepted as being classical. This criterion isn't very strict and dynamics could possibly ruin the classical behaviour of such a state. However, in this work, we weren't concerned about the details of dynamics but just what fraction of randomly chosen states in a sample with large number of states can satisfy our classicality criterion. 

Since it is often assumed that large occupation number of a single mode for a system of a large number of identical Bosons leads to classical field-like behaviour, we tested this assumption. In section \ref{sec:results}, we found that picking states at random with large $\langle N \rangle$ doesn't imply that $\sigma_\varphi <  \frac{\langle \varphi \rangle}{2} $. 
Thus, as is clear from fig (\ref{fig:posneg}), an arbitrary state of quantum field with high occupation number but otherwise arbitrary expansion coefficients (in a basis of number eigenstates), is incredibly unlikely to resemble a classical field (in the sense that quantum fluctuation in field is small compared to its expectation value). As we saw in section \ref{sec:results}, imposing some restrictions on the state vectors in our sample can improve the classical behavior, see fig (\ref{fig:pospos}).

We also looked at large samples of randomly chosen states which deviate from coherent state by an amount which could be chosen \ref{sec:results}. We found that as the states deviate from coherent state, their classical behaviour is also ruined, see fig (\ref{fig:coherent}).
Thus, the proximity of a state to coherent state is a better criterion to decide the validity of classical behaviour rather than its occupation number. 

Before closing, we mention a few issues worth thinking about. 
In hindsight, the best way to interpret our results  is to realise that the classical-like states form a small islands in the vast state space of the quantum  system we are dealing with.
If a system of identical Bosons has $N_{\rm tot}$ particles and there are $M$ oscillator modes, how many distinct quantum states $| \{ N_j \} \rangle$ (of course $N_j \ge 0$) are such that $\sum_{j = 1}^M = N_{\rm tot}$? This number of ways to distribute $N_{\rm tot}$ identical Bosons into $M$ modes is $\binom{N_{\rm tot} + M - 1}{N_{\rm tot}}$ and one less than this is the dimension of the state space of this quantum system of Bosons \cite{Sikivie:2016enz}.
 As is emphasized in \cite{Sikivie:2016enz}, the number of real parameters to specify a classical state of $M$ oscillators is $2M - 1$ which is far smaller than the dimension of the state space of a system of $N$ bosons distributed in $M$ modes. Thus, a classical field is a low-dimensional approximation to a vastly larger quantum Hilbert space. Our results suggest that small deviations from coherent state subspace can cause the quasi-classical behaviour to easily break down.

If states resembling classical field behaviour are this rare, one needs to provide a mechanism to explain why the quantum field happens to be in such a state and why its dynamics continues to keep it in the subset of state space with this property - an example of such a mechanism is quantum decoherence \cite{Schlosshauer:2019ewh}.
In this context, coherent states could be pointer states but this also depends on the coupling of the system to an environment \cite{Zurek:1992mv}. 
Coherent states could also be generated generated from harmonic oscillators by linearly coupling the system to classical sources (e.g. a driven oscillator, see chapter 3 of \cite{Mukhanov:2007zz}).
This second mechanism is closer to what would happen in a quantum optics lab in which states could be prepared.

If it is not guaranteed that large occupation number implies the validity of classical field approximation, when classical field description is used do computations for a model of ultra-light dark matter and make predictions which are tested against observations, since the classical behaviour needs to be justified, one needs additional information than just the particle mass. E.g. what physics causes the state to be classical field-like? E.g. could quantum decoherence due to an environment have played a role in this? Won't this require couplings to other species? 

Thus, the conclusions about wave DM are subject to the choice and details of such mechanisms.
Small mass values of ULDM which are under tension may not be under tension depending on other assumptions e.g. could there be scenarios in which the condensate formation isn't completely successful? This further motivates hybrid model of condensate and particle Dark Matter \cite{Proukakis:2023txk}) and mixed Fuzzy Dark Matter (FDM) - Cold Dark Matter (CDM) simulations \cite{Lague:2023wes}.

We restricted our attention to a single mode since in many cases of interest, only one mode has too large occupation number. One might wish to examine what will happen when we take into account multiple modes.
Similarly, in this work, we stayed focussed on pure states and didn't consider mixed states for the mode of interest of the quantum field.
We could also, in principle, work with other states of interest e.g. squeezed states.

More generally, classical fields are very common not only in cosmology where they are used to model dark energy, wave dark matter, infrared modifications of gravity, screening mechanisms, inflaton etc, but also in other areas of physics e.g. optics, Bose-Einstein condensation etc. While we were concerned about the validity of classical field theory in the context of wave DM, a lot this analysis holds in general.

\begin{acknowledgements}
The author would like to thank Pinaki Majumdar for useful discussions. The work of of the author is supported by Department of Science and Technology, Government of India under Anusandhan National Research Foundation (formerly Science and
Engineering Research Board) - State University Research Excellence (SUR/2022/005391).
\end{acknowledgements}

\end{document}